\documentclass[11pt]{article}
\usepackage{amssymb}
\usepackage{amsmath}
\usepackage{mathtools}
\usepackage[english]{babel}
\usepackage{hyperref}
\usepackage{xcolor}

\newcommand{\be}{\begin{equation}}
\newcommand{\ee}{\end{equation}}
\newcommand{\bea}{\begin{array}}
\newcommand{\ea}{\end{array}}
\newcommand{\beqa}{\begin{eqnarray}}
\newcommand{\eeqa}{\end{eqnarray}}
\newcommand{\bean}{\begin{eqnarray*}}
\newcommand{\eean}{\end{eqnarray*}}

\def\up#1{\leavevmode \raise.16ex\hbox{#1}}

\newcommand{\gapproxeq}{\lower
.7ex\hbox{$\;\stackrel{\textstyle >}{\sim}\;$}}
\newcommand{\lapproxeq}{\lower .7ex\hbox{$\;\stackrel
{\textstyle <}{\sim}\;$}}

\newcounter{appendice}

\def\thebibliography#1{{\bf REFERENCES\markboth
{REFERENCES}{REFERENCES}}\list
{[\arabic{enumi}]}{\settowidth\labelwidth{[#1]}\leftmargin\labelwidth
\advance\leftmargin\labelsep
\usecounter{enumi}}
\def\newblock{\hskip .11em plus .33em minus -.07em}
\sloppy
\sfcode`\.=1000\relax}

\begin{document}

\centerline{\LARGE  Cosmic acceleration in Regge-Teitelboim gravity  }
\vskip .5cm

\centerline{ S. Fabi\footnote{sfabi@ua.edu},  A. Stern\footnote{astern@ua.edu} and Chuang Xu\footnote{cxu24@crimson.ua.edu}}

\vskip 1cm

\begin{center}
  { Department of Physics, University of Alabama,\\ Tuscaloosa,
Alabama 35487, USA\\}

\end{center}

\vskip 0.5cm

\abstract{The Regge-Teitelboim formulation of gravity, which utilizes dynamical embeddings in a background space, effectively introduces source terms in the standard Einstein equations that are not attributable to the energy-momentum tensor.  We show that for a simple class of embeddings of the Robertson-Walker metric, these source terms naturally generate cosmic acceleration.}
\newpage

\section{Introduction}
Long ago, Regge and Teitelboim wrote down a formulation of gravity which relies on  embedding  the space-time manifold in a higher dimensional background, and  regards the embedding coordinates, rather than the metric tensor, as dynamical degrees of freedom, in the spirit of string theory.\cite{Regge:2016gaw}.  The theory was further examined in \cite{Deser:1976qt},\cite{Pavsic:1984kv},\cite{Tapia:1988ze},\cite{Maia:1989du},\cite{Bandos:1996rw},\cite{Capovilla:2006dc}.
In addition to its relation with strings and membranes, the theory shares some similarities to certain matrix models in the continuum limit, where the matrices reduce to continuous variables which define embeddings of continuous manifolds in a background space.\cite{Steinacker:2010rh}
Dynamics in that case is governed by matrix actions (see \cite{Blaschke:2010rg}), while for the Regge-Teitelboim (RT)  formulation of gravity the dynamics is governed by the Einstein-Hilbert action, expressed in terms of embedding coordinates. 

Many  articles have been devoted to cosmological-type solutions to matrix models (see for example \cite{Freedman:2004xg},
\cite{Klammer:2009ku},\cite{Kim:2011cr},\cite{Stern:2014aqa}), while not  much attention has been previously given to  cosmological solutions to RT gravity.   (For examples, see\cite{Davidson:1997ys},\cite{Cordero:2013pua}.)
Further examination of the role of cosmological solutions to RT gravity appears to be warranted.
The RT  formulation of gravity effectively introduces source terms in the standard Einstein equations that are not attributable to the energy-momentum tensor,  suggesting that it is well suited  to as a candidate for a source  for cosmic acceleration. These additional contributions  $T_{{}_{\rm RT}}^{\mu\nu}$ to the Einstein equations are due solely to the embedding of the space-time manifold in the background geometry.  Moreover, $T_{{}_{\rm RT}}^{\mu\nu}$ is strongly dependent on the type of embedding, and a number of constraints restrict its form.  In fact, for some embeddings, the constraints  are too restrictive  to allow for a nonvanishing  $T_{{}_{\rm RT}}^{\mu\nu}$.
The example we consider in this article is the  embedding of the Robertson-Walker metric in a five-dimensional flat background.  (As indicated above, the embedding is not unique.  See for example \cite{Paston:2013uia}.)  The  embedding we choose leads to modifications of the standard Friedmann equations for $k=1$ and $-1$.\footnote{Embeddings of the Robertson-Walker metric into higher dimensional spaces are known to induce corrections  to the Friedmann equations.\cite{Wesson}}
 However, we get no such modification for the case of $k=0$. Our embedding is distinct from the one used   previously by Davidson\cite{Davidson:1997ys}, and leads to a different modification of the standard Friedmann equations.  The system presented in this article  has the advantage that it allows for exact solutions in the case where the usual energy-momentum tensor vanishes.  Numerical solutions are found for the  case where  nonrelativistic matter and radiation, as well as  from  $T_{{}_{\rm RT}}^{\mu\nu}$, are present.  The $k=-1$ solutions describe  an accelerating expansion at large times.   Moreover, solutions exist which exhibit a transition from a decelerating universe to an accelerating one.

We review the  RT formulation of gravity in section 2, and apply it to an embedding of the  Robertson-Walker metric (with $k=\pm 1$) in a  five-dimensional  flat background in section 3.  A modification of the standard Friedmann equations is obtained, which is solved for  various special cases.   Concluding remarks are given in section 4.

\section{Regge-Teitelboim formulation of gravity}

In the Regge-Teitelboim formulation of gravity
 the four-dimensional space-time manifold $M$  is embedded in a flat background  $\mathbb{R}^{m,n}$,  and the   metric tensor $g_{\mu\nu}\,,\; \mu,\nu\cdots=0,...,3,$ on $M$ is   induced from the background metric  $\eta_{ab}\,,\; a,b=0,...,m+n$.  Upon denoting the embedding functions  by $Y^a(x)$,
  where $x^\mu$ are coordinates on $M$, 
\be  g_{\mu\nu}=\eta_{ab}\partial_\mu Y^a\partial_\nu Y^b\;,\ee
 $\partial_\mu$ denoting differentiation with respect to $x^\mu$. 
 From metric compatibility, $\nabla_\lambda g_{\mu\nu}=0$, one gets the conditions\cite{Regge:2016gaw} \be \eta_{ab}\partial_\mu Y^a\nabla_\lambda\partial_\nu Y^b=0\label{mbdng}\ee
The usual Einstein-Hilbert action  \be S=S_{EH}+ S_{source}, \qquad S_{EH}=\frac 1{16\pi G}\int_M d^4x\sqrt{|g|} R\;,\label{actnwthsrs}\ee
governs the dynamics of RT gravity, with the caveat that the dynamical degrees of freedom in this case are the embedding functions, rather than the metric tensor.  Thus $R$ in (\ref{actnwthsrs}) is the Riemann curvature scalar expressed as a function of $ Y^a$.  
 The field equations resulting from extremizing the action with respect to $Y^a$
are
\be \nabla_\mu\Bigl(\left(G^{\mu\nu}-8\pi G\,T^{\mu\nu}\right)\partial_\nu Y^a\Bigr)=0\;,\label{rteom}\ee 
where $G^{\mu\nu}$ is the Einstein tensor, and $T^{\mu\nu}$ is the energy-momentum tensor that arises from the variations of $ S_{source}$. 
  $T^{\mu\nu}$ is covariantly conserved, provided that the metric tensor is nonsingular.  This is a consequence of the conditions (\ref{mbdng}).   Applying the   Bianchi identities to (\ref{rteom}) gives
\be (G^{\mu\nu}-8\pi G T^{\mu\nu})\nabla_\mu\,\partial_\nu Y^a=8\pi G\nabla_\mu T^{\mu\nu} \,\partial_\nu Y^a\ee 
Then after contracting the left- and right-hand sides with $\partial_\lambda Y_a$, and applying  (\ref{mbdng}), one gets
\be \nabla_\mu T^{\mu\nu}  g_{\lambda \nu}=0\ee 

The equations of motion  (\ref{rteom}) can be expressed as current conservation laws on $M$
\be \partial_\mu {\cal J}^{\mu\,a}=0\;,\qquad {\cal J}^{\mu\,a}=\sqrt{|g|} (G^{\mu\nu}-8\pi G\,T^{\mu\nu})\partial_\nu Y^a\;,\label{feon}\ee
  The currents effectively introduce additional source terms in the Einstein equations
\be G^{\mu\nu}=8\pi G\,(T^{\mu\nu}+T_{{}_{\rm RT}}^{\mu\nu})\;,\label{mdfdeeq}\ee
\be   T_{{}_{\rm RT}}^{\mu\nu}=\frac 1{16\pi G\sqrt{|g|} }\eta_{ab}\, ( {\cal J}^{\mu\,a}\partial^\nu Y^b+ {\cal J}^{\nu\,a}\partial^\mu Y^b)\label{RTTmunu}\ee
Obviously, $T_{{}_{\rm RT}}^{\mu\nu}$ is covariantly conserved since  $T^{\mu\nu}$ is, the latter following from the equation of motion (\ref{feon}). 
 On the other hand,  the converse is not necessarily true.  Given  $\nabla_\mu T_{{}_{\rm RT}}^{\mu\nu}=0$, we only have
\be  \partial_\mu {\cal J}^{\mu\,a}=8\pi G\,\sqrt{|g|}\, T_{{}_{\rm RT}}^{\mu\nu}\nabla_\mu \partial_\nu Y^a\;,\label{trfeq}\ee
The right hand side of (\ref{trfeq})  is  required to vanish according to the RT equations of motion, thus constraining  $ T_{{}_{\rm RT}}^{\mu\nu}$.  Symmetry considerations put further restrictions on    $ T_{{}_{\rm RT}}^{\mu\nu}$, which can severely restrict the form of these source terms.   In fact, for many embeddings the constraints do not allow for a nonvanishing  $ T_{{}_{\rm RT}}^{\mu\nu}$.  In the example which follows we examine an embedding of the Robertson-Walker metric, and as a result of the above constraints, the expression for $ T_{{}_{\rm RT}}^{\mu\nu}$ is uniquely determined  up to an overall constant.

\section{Application to Cosmology}
\setcounter{equation}{0}

Starting with the  Robertson-Walker metric
\be ds^2=-dt^2+a(t)^2\Big(\frac {dr^2}{1-kr^2}+r^2(d\theta^2+\sin^2\theta d\phi^2)\Bigr)\;,\ee
describing maximally isometric  $t$-slices, we introduce an energy-momentum tensor $T^{\mu\nu}$, as well as the additional source terms $ T_{{}_{\rm RT}}^{\mu\nu}$.  Before writing down the embedding, we write down the expected modification of the Friedmann equations, and   mention the constraints which restrict the form of  $ T_{{}_{\rm RT}}^{\mu\nu}$.  As usual, we assume that 
 $T^{\mu\nu}$ is associated with a perfect fluid. It is natural to extend  this assumption to the additional source term $ T_{{}_{\rm RT}}^{\mu\nu}$ as well, and further
 that  both $T^{\mu\nu}$ and $ T_{{}_{\rm RT}}^{\mu\nu}$ have a common co-moving frame, as this will preserve the maximal isometry.  Thus 
\beqa T^{00}=\rho(t) &\qquad& T^{ij}=p(t) g^{ij}\label{TttTij}\\&&\cr
 T_{{}_{\rm RT}}^{00}=\rho_{{}_{\rm RT}}(t) &\qquad &T_{{}_{\rm RT}}^{ij}=p_{{}_{\rm RT}}(t) g^{ij}\;,\label{TndTRT}\eeqa 
with all other components vanishing.  $i,j,...$ denote spatial indices, and $0$ is the time index.  As these tensors are required to be covariantly conserved,
\beqa  \dot \rho+3\frac{\dot a}a(\rho+p)&=&0\label{cnsrvtneek}\\&&\cr
 \dot \rho_{{}_{\rm RT}}+3\frac{\dot a}a(\rho_{{}_{\rm RT}}+p_{{}_{\rm RT}})&=&0\label{cntnuitRT}\;,
\eeqa  the dot denoting a $t-$derivative. Substituting (\ref{TttTij}) and (\ref{TndTRT}) into (\ref{mdfdeeq}) leads to the following  Friedmann equations for $a(t)$
\be  \Bigl(\frac{\dot a}a\Bigr)^2=\frac{8\pi G}3(\rho+\rho_{{}_{\rm RT}})-\frac k{a^2}\label{frdmnone}\ee
\be  \frac{\ddot a}a=-\frac{4\pi G}3\Bigl(\rho+\rho_{{}_{\rm RT}}+3( p+ p_{{}_{\rm RT}})\Bigr)\;,\ee

  Given some embedding $Y^a(t,x^i)$, $a=0,1,...{D-1}$ ($x^i$ denoting spatial coordinates) of the space-time manifold in a $D-$dimensional background, the currents $ {\cal J}^{\mu\,a}$ in  (\ref{feon}) have the form
\beqa  {\cal J}^{0\,a}&=& 8\pi G\,\sqrt{|g|}\,\rho_{{}_{\rm RT}}(t) \partial_0 Y^a\cr&&\cr
{\cal J}^{i\,a}&=& 8\pi G\,\sqrt{|g|}\,p_{{}_{\rm RT}}(t) g^{ij}\partial_j Y^a\;,\eeqa
The requirement of current conservation leads to the following  constraints on $\rho_{{}_{\rm RT}}(t)$ and $p_{{}_{\rm RT}}(t)$
\be  \rho_{{}_{\rm RT}}(t) \nabla_0 \partial_0 Y^a+ p_{{}_{\rm RT}}(t)g^{ij} \nabla_i \partial_j Y^a=0\;, \label{cnnrap}\ee
 which  depends on the choice of the embedding.  

Next we shall pick an embedding and this will lead to a particular form for  $\rho_{{}_{\rm RT}}(t)$ and $p_{{}_{\rm RT}}(t)$.
We consider the embedding of the Robertson-Walker metric  with $ k=\pm1$ in a five-dimensional  flat background, and make the following choice\cite{JRosen}\footnote{The choice is not unique.  Another embedding in five dimensions is applied in  \cite{Davidson:1997ys}, which leads to a different modification of the Friedmann equations.}
\be  \begin{pmatrix}
Y^0 \\Y^1 \\Y^2\\Y^3\\Y^4
\end{pmatrix}
=\begin{pmatrix}b_k(t)\\ a(t) r\sin\theta\sin\phi\\ a(t) r\sin\theta\cos\phi \\ a(t)r\cos\theta\\ a(t) \sqrt{1-k r^2}\end{pmatrix}\;,\label{mbdng2}\ee
where $0< r<1$ for $k=1$ and $0< r<\infty$ for $k=-1$, along with $0\le\theta\le \pi$, $0\le\phi<\pi$.  For the background metric take $(-,+,+,+,k)$.  The background space is therefore $\mathbb{R}^{4,1}$ for $k=1$, and 
 $\mathbb{R}^{3,2}$ for $k=-1$. In order to recover the Robertson-Walker induced metric  one needs to impose  that $\dot b_k=\sqrt{1+k\dot a^2}$, making $b_k(t)$ nonlocal in  time. 
Upon substituting directly into the equations of motion (\ref{rteom}), and assuming that the stress-energy tensor has the form (\ref{TttTij}), one gets 
\be k\dot a\dot b_k ^4  +3a \dot a \ddot a
\dot b_k^2  =\frac{8\pi G}3  a^2\left( ka \dot a \ddot a
    \rho+  a\dot b_k^2  \dot \rho+3  \dot a \dot b_k^2 \rho \right)\;,\label{Y4keqwn}\ee
along with the conservation equation (\ref{cnsrvtneek}).  Integration of  (\ref{Y4keqwn}) gives
\be ka\dot b_k^3=\frac{8\pi G}3  ( a^3 \dot b_k
     \rho+ c_0)\;,\ee
where $c_0$ is an integration constant.  It can be rewritten as
\be \frac{\dot a^2}{a^2}=\frac{8\pi G}3 \Bigl(
     \rho+ \frac{c_0}{  a^3\sqrt{1+k\dot a^2}}\Bigr)-\frac{k}{a^2}\;,\label{mdfdFwq}\ee
We thus recover the standard Friedmann dynamics when $c_0=0$. More generally,   upon comparing (\ref{mdfdFwq}) with (\ref{frdmnone}) we get that 
\be \rho_{{}_{\rm RT}}=
    \frac{c_0 }{  a^3\sqrt{1+k\dot a^2}}\label{rhort} \ee
$p_{{}_{\rm RT}}$ can then be obtained from the conservation equation (\ref{cntnuitRT})
\be p_{{}_{\rm RT}}=-\frac a{3\dot a}\dot\rho_{{}_{\rm RT}}-\rho_{{}_{\rm RT}}=\frac{k c_0 \ddot a}{3 a^2 \left(1+k\dot a^2\right)^{3/2}}\label{pressurert}\ee Combining (\ref{rhort}) and (\ref{pressurert}) gives the analogue of an equation of state for $T_{{}_{\rm RT}}^{\mu\nu}$:
\be p_{{}_{\rm RT}}=\frac {k a \ddot a}{3 (1+k\dot a^2)}\rho_{{}_{\rm RT}}\label{eqsttrt}\ee

The quantity in parenthesis in (\ref{mdfdFwq}), times $a^2$, can be regarded as minus an effective potential, $V_{eff}(a)$. For this to make sense,  the second term in parenthesis should  be evaluated on-shell, thereby making $\dot a $ a function of $a$. It  gives the following contribution to the   effective force  $-\frac d{da}V_{eff}$:
\be -\frac{8\pi Gc_0}{3  a^2(1+k\dot a^2)^{3/2}}\Big[1+k\dot a^2+k a \dot a \frac{d\dot a}{da}\Big]\label{drkfrc}\ee
This can be considered to be  a `dark force', in the sense that it does not originate from the energy momentum tensor.
Provided that the quantity in brackets is positive, it  follows that  this is  an attractive force when $c_0 >0$, and  a repulsive force when  $c_0<0$.

It is usual to relate  the spatial geometry, i.e., $k$, to the density parameter $\Omega=\rho/\rho_c$,  $\rho_c=3H^2/8\pi G$ being the critical density.
From eq. (\ref{mdfdFwq}) we get
\be \frac{k}{a^2}= 
     H^2\Bigl(\Omega- 1+ \frac{{\tilde c_0}}{H^2  a^3\sqrt{1+k a^2H^2}}\Bigr)\;,\ee
where  $H=\dot a/a$  is the  Hubble parameter and ${\tilde c_0}=\frac{8\pi Gc_0}3$. 
When ${\tilde c_0}$  is  zero, we recover the usual result, that $k=-1,1$ corresponds to $\Omega<1,\Omega>1$, respectively. [Recall, $k=0$ is not valid for the embedding (\ref{mbdng2}).]   We can make the same argument when ${\tilde c_0}\ne 0$, provided we replace the density parameter  $\Omega$ by 
\be\Omega'=\Omega+ \frac{{\tilde c_0}}{H^2  a^3\sqrt{1+k a^2H^2}}\label{dnstee}\ee 

Let us assume the standard  equation of state relation for $p$ and $\rho$
\be p=w\rho\;\ee 
Then from  (\ref{cnsrvtneek}) it follows that 
$ \frac d{da}(a^3\rho)=-3wa^2 \rho$, and one obtains the usual relation between the density and the scale factor
\be\rho=c_1 a^{-3(w+1)}\;,\ee
$c_1$ being another integration constant.   (\ref{mdfdFwq})
then becomes
 \be{k a} (1+ k{\dot a^2})=\frac{\tilde c_1}{ a^{3w}}
     + \frac{\tilde c_0}{ \sqrt{1+k\dot a^2}}\;,\label{dnsfea}\ee
where  ${\tilde c_m}=\frac{8\pi G}3c_m$, $m=0,1$, from which we  obtain the evolution of the scale factor. 
The first term on the right hand side represents the usual matter/energy contribution, while the second term is a result of the embedding.  For $k=-1$, the second term in (\ref{dnsfea}) is significant when $\dot a$ approaches $1$. Below we examine the  $k=-1$ solutions  for three cases:  the absence of  a normal matter/energy contribution discussed in 3.1,   a nonrelativistic matter component and  a radiation component, discussed, respectively, in 3.2 and 3.3. Exact solutions are found for  3.1, while numerical solutions can be obtained  for   3.2  and  3.3.  Due to the square root in the second term on the right hand side of (\ref{dnsfea}), we get the restriction that $\dot a^2<1$ when $k=-1$ (at least, if we restrict to real values of $\tilde c_0$), and so an  inflationary phase scenario  for $k=-1$ appears  to be problematic in this model.

\subsection{$\tilde c_1=0$}
 In the absence of a matter/energy source,  (\ref{dnsfea}) reduces to
 \be  k{\dot a^2}=  \Bigl(\frac{k\tilde c_0}{ a}\Bigr)^{2/3}-1\;,\label{scleqc1z}\ee  where we need to restrict $\tilde c_0>0$ for $k=1$, and $\tilde c_0<0$ for $k=-1$ in order for $a(t)$ to be real.
The acceleration of the scale factor in this case is simply
\be \ddot a =-\frac{k(k\tilde c_0)^{2/3}}{3 a^{5/3}}\;,\label{xclrtnaz}\ee
which  tends to zero for large $a$.   It is negative for $k=1$ and  positive for $k=-1$.  Substitution of (\ref{scleqc1z}) and (\ref{xclrtnaz})
into (\ref{eqsttrt})   leads to a fixed equation of state for $T_{{}_{\rm RT}}^{\mu\nu}$: 
 $  p_{{}_{\rm RT}}=-\frac 19\rho_{{}_{\rm RT}}$.  (This result was recovered in a certain limit in \cite{Davidson:1997ys}.)   Note that for $k=-1$, $p_{{}_{\rm RT}}>0$ and $\rho_{{}_{\rm RT}}<0$. 

Integration of (\ref{scleqc1z})  gives
\beqa t&=&2\tilde c_0-a\Biggl(2\Bigl(\frac{\tilde c_0}{ a}\Bigr)^{2/3}+1\Biggr)\sqrt{\Bigl(\frac{\tilde c_0}{ a}\Bigr)^{2/3}-1}\;,\qquad\; 0\le t\le 2\tilde c_0\;,\qquad k=1\cr&&\cr
t&=&a\Biggl(2\Bigl(\frac{-\tilde c_0}{ a}\Bigr)^{2/3}+1\Biggr)\sqrt{1-\Bigl(\frac{-\tilde c_0}{ a}\Bigr)^{2/3}}\;,\qquad\quad\;\, t\ge 0\;,\qquad\qquad k=-1\;,\cr&&\eeqa
where we fixed the time origin in both cases in order for $a$ to be a minimum at $t=0$.  The results are plotted in figure 1.
As usual,  $k=1$  and $k=-1$ correspond, respectively, to  closed and open universes. The   $k=1$ solution yields  a singularity  at $t=0$, and $a$ reaches its maximum value of  $|\tilde c_0|$ at $t=2|\tilde c_0|$. For  the  $k=-1$ solution, $a$ has a nonzero minimum value  equal to  $|\tilde c_0|$ at $t=0$, and its  continuation to $t<0$ describes a bouncing cosmology.  In this case,  $ \dot a$ is zero at $t=0$, while  it approaches a constant value at large $t$, which is consistent with an accelerating cosmology.  We will focus on accelerating cosmologies in the following two cases and hence set $k=-1$.
  \begin{figure}[h]
\begin{center}
\includegraphics[height=2in,width=3in,angle=0]{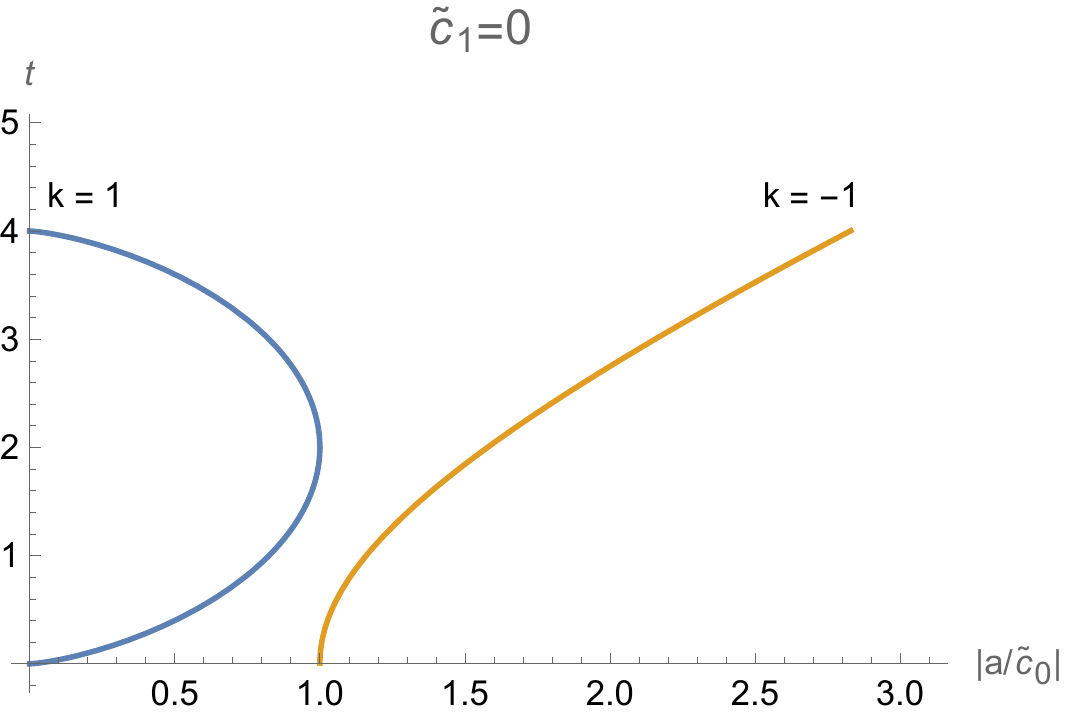}
\caption {Plot of $ t$ as a function of $|a/\tilde c_0| $ when $\tilde c_1=0$  for the cases $k=1$ and  $k=-1$.  }
\end{center}
\end{figure}

\bigskip

\subsection{$w=0$  and $k=-1$}
This case includes  a  nonrelativistic matter component for $k=-1$, in addition to the RT source.  Perturbations  around the previous  solution  with $\tilde c_1=0,\;k=-1$ are obtained for $c_0$  real and negative.  Furthermore, eq. (\ref{dnsfea}) allows for the  initial conditions $\dot a=0$ and $\ddot a>0$ at $t=0$ provided that  $-\tilde c_0-\tilde c_1= a(0)\ge 0$. The minimum value of the scale parameter, $ a(0)$, can therefore be nonzero when  $-\tilde c_0>\tilde c_1$, again describing a bouncing cosmology.  An example, $\tilde c_0=-\frac 23,\tilde c_1=\frac 13$,  is  plotted in figure 2, and it has $\ddot a>0$. The ratio $p_{{}_{\rm RT}}/\rho_{{}_{\rm RT}}$, evaluated on-shell, is not a constant for this case. 

 The previous solution, $\tilde c_0=-1\,,\tilde c_1=0$,  as well as  $\tilde c_0=0,\tilde c_1=1$ (corresponding to the absence of the RT source term), are  shown in figure 2 for comparison purposes.  The $\tilde c_0=0,\tilde c_1=1$ solution has $\dot a>1$, and so, from eq. (\ref{dnsfea}), in order to perturb around that solution one would need that $\tilde c_0$ is purely imaginary.  We find that these perturbations do not alter the  $\tilde c_0=0,\tilde c_1=1$  solution significantly, and moreover   do not appear to exhibit a positive acceleration. 
 \begin{figure}[h]
\begin{center}
\includegraphics[height=2.85in,width=3.75in,angle=0]{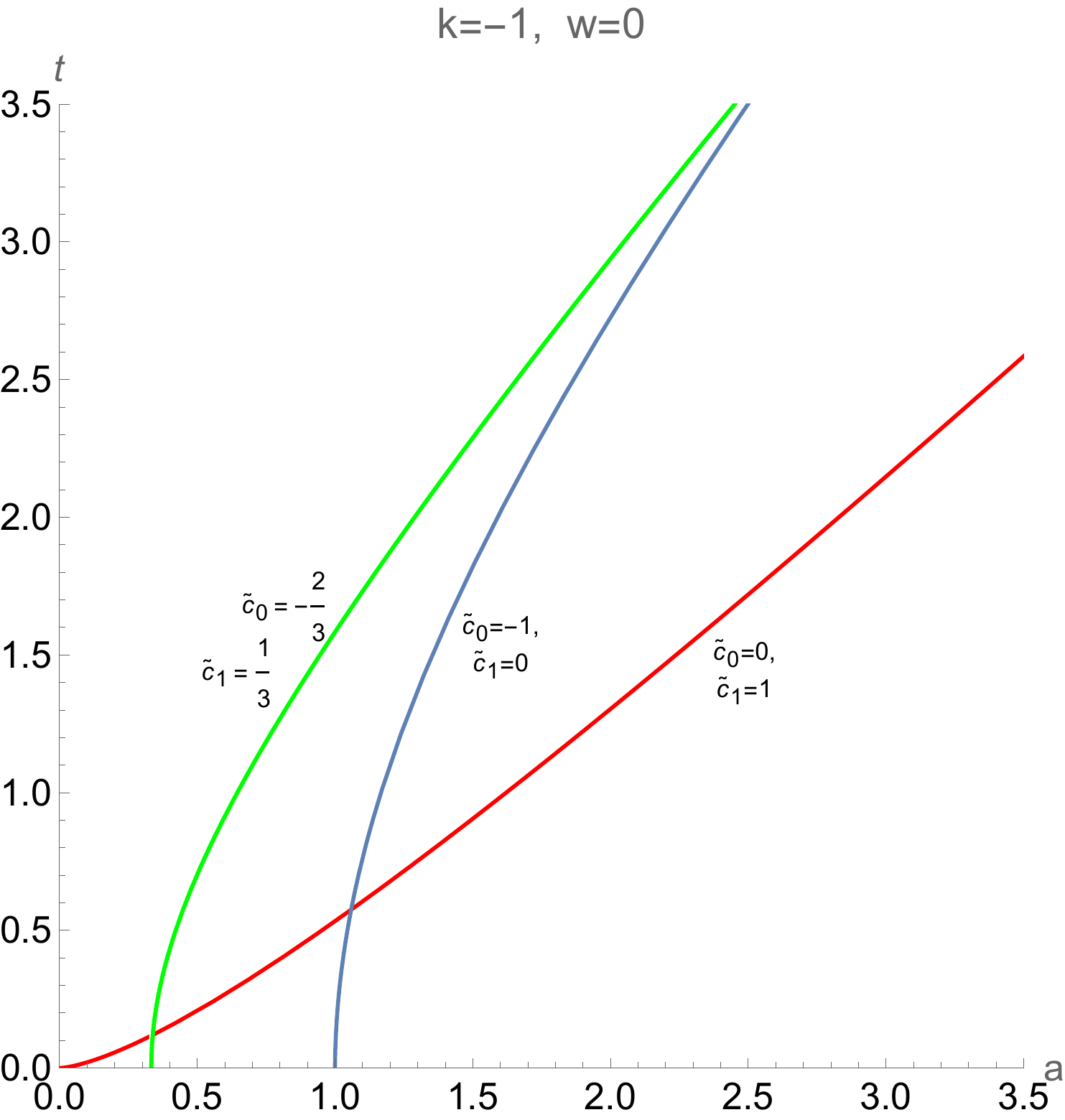}
\caption {Plot of $ t$ as a function of $a $   for   $k=-1$, $w=0$ for the cases: \newline
 $\;i)\;\tilde c_0=-1,\tilde c_1=0\;$,
  $\;ii)\;\tilde c_0=-\frac 23,\tilde c_1=\frac 13\;$  and  $\;iii)\;\tilde c_0=0,\tilde c_1=1$.  }
\end{center}
\end{figure}

\bigskip

\subsection{$w=\frac 13$  and $k=-1$}
This case includes  a  radiation  component  for $k=-1$, along with the Regge-Teitelboim source.  Once again, perturbations  around the  solution  with $\tilde c_1=0,\;k=-1$ are obtained for $c_0$  real and negative. Two kinds of initial conditions can be chosen:  1) $\dot a=0$ and $\ddot a>0$ at $t=0$ or  2)  $a(0)=0$.  From  eq. (\ref{dnsfea}), 1) is possible when  $\frac 12(-\tilde c_0\pm\sqrt{\tilde c_0^2-4\tilde c_1})= a(0)\ge 0$.  $a(0)$ then corresponds to a minimum, leading  to a bouncing cosmology.   2) describes a singularity.
Two nontrivial solutions  are plotted in figure 3 $\;i)\;\tilde c_0=-.85,\tilde c_1=.15\;$,  and  $\;ii)\;\tilde c_0=-.8,\tilde c_1=.2\;$ 
  $a$ has a nonzero minimum value for the former, while $a(0)=0$ for 
the latter. $\ddot a$ is positive for all $t$ for $\;i)$.  On the other hand, for case  $\;ii)$, starting from a value of $\dot a\approx 1$ in the limit $t\rightarrow 0$, the system initially decelerates, and then makes  a transition  to an accelerating phase  ($\ddot a>0$) at a  finite time $t$.  More generally, from (\ref{dnsfea}), such a transition occurs when $\dot a^2+\frac{\tilde c_1}{a^2}=1$.

Plots for $\;iii)\;\tilde c_0=-1,\tilde c_1=0\;$,
   and  $\;iv)\;\tilde c_0=0,\tilde c_1=1$ are included in figure 3 for comparison purposes. As in the previous subsection,  the $\tilde c_0=0,\tilde c_1=1$ solution has $\dot a>1$, and so, from eq. (\ref{dnsfea}),  perturbations around that solution are  not possible for real values of  $\tilde c_0$. 
 \begin{figure}[h]
\begin{center}
\includegraphics[height=2.85in,width=3.75in,angle=0]{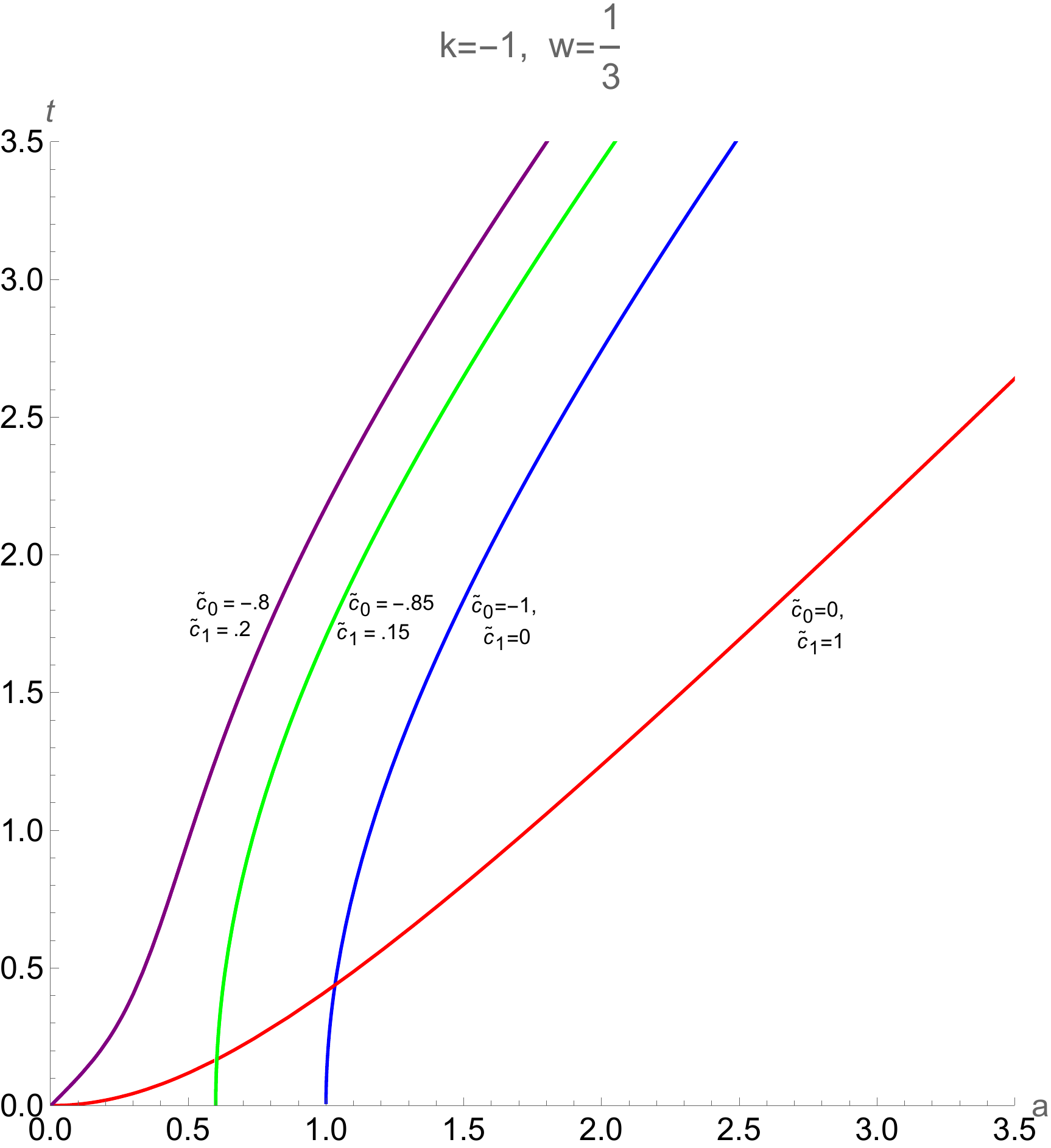}
\caption {Plot of $ t$ as a function of $a $   for   $k=-1$, $w=\frac13$ for the cases: \newline
   $\;i)\;\tilde c_0=-.85,\tilde c_1=.15\;$,   $\;ii)\;\tilde c_0=-.8,\tilde c_1=.2\;,$ $\;iii)\;\tilde c_0=-1,\tilde c_1=0\;$, and  $\;iv)\;\tilde c_0=0,\tilde c_1=1$.  }
\end{center}
\end{figure}

\newpage
\section{ Concluding remarks}
The introduction of the RT source term in the Einstein equations leads to the modification of the Friedmann equations (\ref{dnsfea}). The modification can be interpreted as  a `dark force', as it does not originate from the energy-momentum tensor.  For the choice of embedding used in this paper, this force was given by (\ref{drkfrc}).   The modification of the Friedmann equations also gives an additional contribution to the density parameter (\ref{dnstee}).
Exact solutions were found for the case where the usual energy-momentum tensor vanishes, which describe  an accelerating universe when $k=-1$.   This result persisted when  a nonrelativistic matter component or  radiation component was included.  Moreover, for the case of  a  radiation component, there are $k=-1$ solutions which undergo  a transition from a decelerating universe to an accelerating one.

Our results rely on a particular choice for the embedding\cite{JRosen}.  It remains to be seen how robust the results are with respect to other choices of  embeddings.  Different `dark' source terms  $T_{{}_{\rm RT}}^{\mu\nu}$ will result from different choices of the embedding.  These source terms are highly constrained by  the current conservation conditions in (\ref{trfeq}), and by symmetry. For the case of the Robertson-Walker metric, with  sources having the form in (\ref{TndTRT}), the constraints are given by  (\ref{cnnrap}).  For some embeddings, the conditions are too strong to allow for any 
 modification of the Einstein equations.  This is the case for the embedding  of the $k=0$ Robertson-Walker metric in \cite{JRosen}.  As $k=0$ is of physical relevance, it is worthwhile to do an extensive  search for  embeddings that produce 
analogous `dark' contributions to the Friedmann equations.  Another possible extension of this work  is to relax the requirement that the  background space is geometrically flat,  and perhaps only demanding that it be Ricci flat.\cite{Wesson},\cite{Romero:1996ba}  Of course, it is of interest to work towards constructing a more realistic cosmological model, one for example, that includes different matter-energy components, and  combines  the radiation and nonrelativistic matter eras.
We plan to address these and other issues in the future.

\bigskip
{\bf Acknowledgment} 

A.S. is grateful for discussions with A. Pinzul.

\end{document}